\documentclass[tran]{IEEEtran}
\usepackage{graphicx,subcaption,float}
\usepackage[numbers]{natbib}
\usepackage{amssymb}
\usepackage{amsmath}
\usepackage{hyperref}
\usepackage{soul}
\usepackage{booktabs}
\usepackage[boldmath]{numprint}
\npdecimalsign{.}
\nprounddigits{3}

\usepackage[nomargin,inline,final]{fixme}
\fxusetheme{color}
\FXRegisterAuthor{ra}{ara}{roberto}
\graphicspath{{imgs/}}

\begin{document}

\title{Multi-feature 360 Video Quality Estimation}

\author{
  Roberto~G.~de~A.~Azevedo, 
  Neil~Birkbeck, 
  Ivan~Janatra, 
  Balu~Adsumilli, 
  Pascal~Frossard 
\thanks{R. Azevedo and P. Frossard are with
        Signal Processing Lab.~(LTS4),
        École Polytechnique Fédérale de Lausanne~(EPFL),
        Lausanne, Switzerland.}%
\thanks{N. Birkbeck, I. Janatra, and B. Adsumilli are with
        Youtube,
        Mountain View, California, USA.}}

\maketitle

\date{Received data: \today}


\begin{abstract}
We propose a new method for the visual quality assessment of
360-degree~(omnidirectional) videos.
The proposed method is based on computing multiple spatio-temporal objective
quality features on viewports extracted from 360-degree videos.
A new model is learnt to properly combine these features into a metric
that closely matches subjective quality scores.
The main motivations for the proposed approach are that:
1)~quality metrics computed on viewports better captures the user experience
than metrics computed on the projection domain;
2)~the use of viewports easily supports different projection methods being
used in current 360-degree video systems; and
3)~no individual objective image quality metric always performs the best for
all types of visual distortions, while a learned combination of them is able
to adapt to different conditions.
Experimental results, based on both the largest available 360-degree videos
quality dataset and a cross-dataset validation, demonstrate that the proposed
metric outperforms state-of-the-art 360-degree and 2D video quality metrics.
\end{abstract}


\maketitle


\ranote{Overall warning:  Still need a lot of clean-up and improve the results
section.
What to add from the ICME paper: 1. Experiments in the whole VQA-ODV dataset.
2. Maybe also our dataset?
3. Ablation study (including importance of the different viewports?  Can we
show the most used viewports of our model are the same being most viewed by
the users?
4. Weighted-version of the metrics based on VA.
Also possible: NR individual metrics?
}

\section{Introduction}
Driven by the growing interest in virtual and augmented reality,
omnidirectional~(or 360-degree) videos are becoming prevalent in many
immersive applications, e.g., medicine, education, and entertainment.
Omnidirectional videos are spherical signals captured by cameras with a full
360-degree field-of-view~(FoV).
When consumed via head-mounted displays~(HMDs) omnidirectional videos allow
the user to be immersed in the content.
During runtime, based on the user's head motion, the portion of the sphere in
the user's field of view, named \emph{viewport}, is seamlessly updated
following the user's head motion, thus providing an improved sense of
presence.
Together, the new immersive features and interactive dimension change the end user perceived quality of experience~(QoE) in many ways when
compared to traditional videos~\cite{azevedo_visual_2020}.
Similarly to traditional audiovisual multimedia content, methods for assessing
the QoE of omnidirectional content plays a central role in shaping processing
algorithms and systems, as well as their implementation, optimization, and
testing~\cite{lin_perceptual_2011}.
In particular, the visual quality assessment of omnidirectional videos is one
of the most important aspects of users' QoE when consuming such immersive
content.

Quality assessment of 360-degree visual content consumed through HMDs brings
its own specificities.
For instance, to reuse existing image and video processing technologies, the
360-degree visual content is commonly mapped to a 2D plane~(the projection
domain) and stored as a rectangular
image~\cite{azevedo_visual_2020,li_state-art_2019}.
Examples of commonly used projections include: equirectangular~(ERP),
truncated square pyramid~(TSP), cube map~(CMP), and equiangular cube
map~(EAC)~\cite{chen2018recent}.
The coupled interaction between projection and compression of the resulting
rectangular images, however, brings new types of visual
distortions~\cite{azevedo_visual_2020}.
Also, the magnification of the content, the supported increased field-of-view,
the fact that the user is completely immersed, and the new interactive
dimension, all contribute to the overall perceived visual quality
and QoE~\cite{azevedo_visual_2020}.
Such new features call for the development of new methods and good practices
related to both subjective and objective quality assessment of 360-degree
visual
content~\cite{desimone_omnidirectional_2017,azevedo_visual_2020,li_state-art_2019}.

Subjective video quality assessment~(VQA) methods collect quality judgments
from human viewers through psychophysical experiments.
Subjective VQA has the advantage of being more reliable, since humans are the
ultimate receivers of the multimedia content.
Subjective VQA, however, is expensive, time-consuming, and not suitable for
real-time processing quality control.
Thus, objective VQA algorithms are required to estimate video quality
automatically.
Based on the amount of the reference content they use, objective VQA metrics
can be divided into:
full-reference~(FR) methods, which require complete access to the reference
video;
reduced-reference~(RR) methods, which do not require the complete reference,
but some features that characterize the reference video; and
no-reference~(NR) methods, which do not require any information about the
reference video.
With regards to the prediction accuracy, FR methods are in general more
accurate and more widely applied.
This paper proposes a new FR method for 360-degree VQA; thus the metrics
discussed hereafter are FR schemes.

PSNR-based objective image quality assessment~(IQA) metrics that take into
account the properties of 360-degree images have been recently proposed in the
literature, e.g., S-PSNR~\cite{yu_framework_2015}
CPP-PSNR~\cite{zakharchenko_quality_2016},
and WS-PSNR~\cite{sun_weighted-spherically-uniform_2017}.
Those methods are easy to implement and can be efficiently integrated into
video coders, but their correlation with subjective judgements are far from
satisfactory.
Moreover, when used for video quality assessment they lack a proper modelling
of the temporal characteristics of the human-visual system~(HVS).
VQA methods must also consider the temporal factors apart from the spatial
ones and the contribution and their interaction to the overall video quality.
Therefore, more perceptually-oriented metrics are still required for
360-degree VQA.

In contrast to previous work, we propose a viewport-based multi-metric
fusion~(MMF) approach for 360-VQA.
The proposed approach extends~\cite{azevedo2020viewport}%
\footnote{Compared to~\cite{azevedo2020viewport}, we extend
the individual features used by our model, propose an adapted temporal pooling
method, and provide an extensive new set of experiments, including different
regression methods and a cross-dataset validation.}
and is based on:
i)~extracting spatio-temporal quality features~(i.e., computing
objective IQA metrics) from viewports;
ii)~temporally pooling them taking the characteristics of the human-visual
system~(HVS) into consideration, and;
iii)~then training a regression model to predict the 360-degree video quality.

On one hand, working with viewports allows us to better account for the
final viewed content and naturally supports different
projections~\cite{birkbeck_quantitative_2017,azevedo_subjective_2020}.
On the other hand, the use of multiple objective metrics computed on these
viewports allows our method to have a good performance for the complex and
diverse nature of visual distortions appearing in 360-degree videos.
Indeed, previous work in both traditional
2D~\cite{
liu_image_2013,
rassool_vmaf_2017,rousselot_quality_2019} and
360-degree~\cite{azevedo_subjective_2020} VQA have recognized that even with
the multitude of available objective IQA metrics, there is no single one that
always performs best for all types of distortions.
The combination of multiple metrics is thus a promising approach that can take
advantages of the power of individual metrics to properly estimate 360 video
quality~\cite{liu_multi-metric_2011,rousselot_quality_2019}.

Experimental results, based on the largest publicly available
360-degree video quality dataset, VQA-ODV~\cite{li_bridge_2018}, and on the
VR-VQA48~\cite{xu_assessing_2018} dataset show the viability of our proposal,
which outperforms state-of-the-art methods for 360-degree VQA.

The rest of the paper is organized as follows.
Section~\ref{sec:related_work} presents the related work.
Section~\ref{sec:proposed_method} describes our proposal and highlights the
main contributions of our proposal.
Section~\ref{sec:experimental_setup} presents the experimental setup used to
validate our approach and the experimental results.
Section~\ref{sec:ablation_studies} provides further experiments
through ablations studies.
Finally, Section~\ref{sec:conclusion} brings our conclusions and future work.

\section{Related work}
\label{sec:related_work}

\subsection{Traditional IQA/VQA methods}
\label{subsec:related_work_traditional}

Traditional 2D image and video quality assessment have attracted a lot of
attention in recent
years~\cite{lin_perceptual_2011,bhattacharyya_full_2018}.
Some objective IQA  metrics are well-established today, e.g., Peak
Signal-to-Noise Ratio~(PSNR),
Structural Similarity (SSIM)~\cite{wang_image_2004},
Multiscale Structural Similarity (MS-SSIM)~\cite{wang_multiscale_2003},
Most Apparent Distortion (MAD)~\cite{chandler_most_2010},
Feature Similarity (FSIM)~\cite{zhang_fsim_2011},
Gradient Magnitude Similarity Deviation  (GMSD)~\cite{xue_gradient_2014}, and
Haar wavelet-based Perceptual Similarity
Index~(HaarPSI)~\cite{reisenhofer_haar_2018}.
Different from IQA metrics, VQA metrics must also take into consideration the
temporal dimension and model the temporal aspects of the HVS.
A straightforward and convenient approach to VQA is to use an IQA method on a
frame-by-frame basis and then compute the global quality score by simple
average of Minkowski summation.
However, such approaches do not correctly model the temporal dimension.

Examples of metrics specifically developed for VQA include
VQM~\cite{pinson_new_2004},
MOVIE~\cite{seshadrinathan_motion-tuned_2010},
Vis3~\cite{vu_vis3_2014}, and
SSTS-GMSD~\cite{creutzburg_video_2015}.
\ranote{An overview of the methods?}
Although such metrics have produced interesting results, currently the best
results are achieved by MMF approaches, from which VMAF is one of the most
prominent examples.
VMAF~\cite{rassool_vmaf_2017}\footnote{\url{https://github.com/Netflix/vmaf}}
is an MMF approach for traditional visual content.
The objective metrics used as features to train an SVM~(Support vector
machine) regressor are:
Visual Information Fidelity  (VIF)~\cite{sheikh_image_2006},
Detail Loss Metric~(DLM)~\cite{li2011image},
Anti-Noise Signal-to-Noise Ratio~(AN-SNR), and
Mean Co-located Pixel Difference~(MCPD).
MCPD is used as a simple metric for the temporal dimension.
The SVM-based regressor is trained to provide an output in the range of 0--100
per video frame.
By default, VMAF final output is the average of the individual frame VMAF
scores, which is clearly not the best approach for modeling the temporal
aspects of the HVS~\cite{sampath2019block-based}.



In this paper, similar to VMAF, we also propose an MMF approach, but we focus
on omnidirectional visual content and take into consideration the perceptual
particularities of this new media type.
In particular, we compute the individual features in the viewports domain,
which allows our method to better account for the final viewed content.
Our method uses a different set of individual spatial and temporal features
and pooling method~(detailed in Section~\ref{sec:proposed_method}) which
support a better correlation to subjective tests.
Finally, in our proposal we use a Random Forest Regression
model~(RFR)~\cite{breiman2001random} whereas VMAF uses SVR.
Such choices are supported by the experimental results in
Section~\ref{sec:experimental_setup}.

\begin{figure*}[t]
  \centering
  \includegraphics[width=0.90\textwidth]{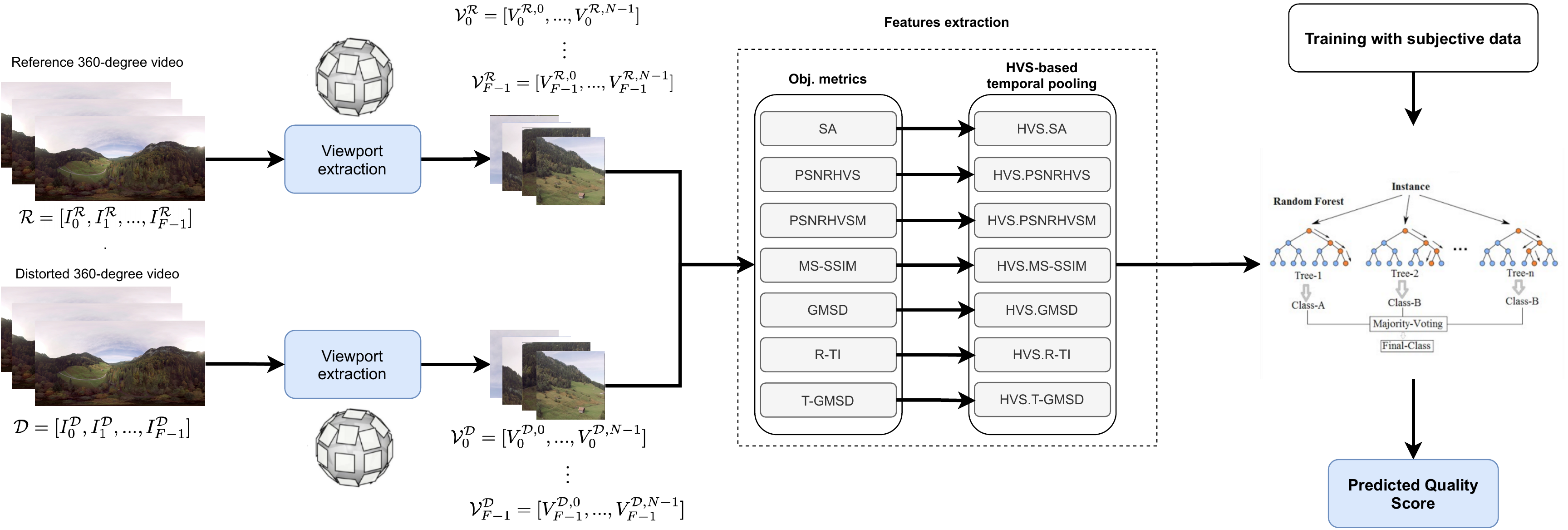}
  \caption{Proposed 360-VQA multi-method approach overview.}
  \label{fig:proposed_method}
\end{figure*}

\subsection{Objective metrics for 360-degree VQA}
Currently, the main approaches for objectively assessing the quality of
360-degree content can be broken into 4 categories:
1)~well-known objective metrics for 2D content computed on the projection
domain;
2)~well-known objective metrics for 2D content computed on the viewports;
3)~objective metrics specifically developed for 360-degree visual content;
and
4)~deep learning techniques.

The use of standard 2D image and video metrics~(e.g., the ones discussed in
Section~\ref{subsec:related_work_traditional}) directly in the projection
domain is straightforward, but they do not properly model the perceived
quality of the 360-degree content.
The main issues with such an approach are two-fold:
first, it gives the same importance to the different parts of the spherical
signal, which not only is sampled very differently from classical images, but
also have different viewing probabilities~(thus different importance);
and second, even for traditional images, these metrics are known to have
limitations for different visual distortion types---none is universally
satisfactory~\cite{liu_image_2013}.
In this paper, we address such issues by computing objective metrics in the
viewport domain and by employing a MMF fusion approach.

To cope with the sampling issue of the projection domain, recent proposals for
omnidirectional image quality assessment have been developed to tackle the
specific geometry of 360-degree images:
Spherical-PSNR~(S-PSNR)~\cite{yu_framework_2015},
Craster Parabolic Projection PSNR~(CPP-PSNR)~\cite{zakharchenko_quality_2016},
Weighted-to-Spherical-PSNR~(WS-PSNR)~\cite{sun_weighted-spherically-uniform_2017},
and S-SSIM~\cite{chen2018spherical}.
In S-PSNR, sampling points uniformly distributed on a spherical surface are
re-projected to the original and distorted images respectively to find the
corresponding pixels, followed by the PSNR calculation.
In CPP-PSNR, the PNSR is computed between samples in the CPP
domain~\cite{zakharchenko_quality_2016}, where pixel distribution is closer
to that in the spherical domain.
The pixels of the original and distorted content are first projected to the
spherical domain and then mapped to the CPP domain, where PSNR is computed.
In WS-PSNR, the PSNR computation at each sample position is performed directly
on the planar domain, but its value is weighted by the area on the sphere
covered by the given sample.
Different weight patterns may be used for different projections.
S-SSIM is a similar approach to S-PSNR, but using SSIM instead of
PSNR~\cite{chen2018spherical}.

The use of objective metrics computed on the viewports is an interesting
approach, in which $N$ viewports of different viewing directions are
generated for both the original and the distorted content, and the 2D metric
is computed individually for each of these viewports.
Then, the overall 360-degree quality metric can be computed by aggregating the
quality of individual viewports.
If the objective metric properly models the human perceptibility, in theory, it
could be a good approximation of the overall 360-degree quality.
The use of viewports~\cite{birkbeck_quantitative_2017,azevedo_subjective_2020}
and Voronoi patches~\cite{croci_voronoi_2019,croci_visual_2020} for computing
individual IQA metrics have also been discussed.
Here, we acknowledge that the use of viewports (or patches) to compute the
360-VQA is indeed a more perceptually-correct ways of assessing 360 visual
quality.
Previous methods, however, simply compute the quality of 360 images as the
average of the viewports~(or patches).
In contrast, we use an MMF approach that allows to better account for different
visual distortion types, the temporal dimension, and viewing probability of
360-degree videos.


Recent works have also proposed deep learning architectures to estimate
360-degree video quality~\cite{li_bridge_2018,kim_deep_2019,li_viewport_2019}.
One of the main issues with such approaches is that the current 360-VQA
datasets are not big enough to satisfactorily train deep learning methods.
Thus, they need to perform data augmentation, such as splitting the original
image into patches or rotating the original 360-degree images.
In both cases, however, it is not clear if the new generated patches or
rotated images share the same quality scores as the original content.

Finally, all the metrics proposed for 360-VQA mentioned above do not
explicitly model the temporal dimension of 360-degree videos. They usually
compute the overall quality simply as the average of the quality of each
individual frame. Different from IQA, however, VQA metrics shall ideally take
the temporal dimension into consideration and properly integrate the temporal
properties of the HVS.
As previously mentioned, some traditional objective VQA metrics do consider the
temporal dimension but do not take into consideration the characteristics of
360-degree videos.

We address the above mentioned issues by computing per-frame spatio-temporal
objective metrics in the viewport domain, temporally pooling them by taking
into account the HVS, and employing a multi-metric fusion approach that
closely matches subjective scores.
As previously mentioned, being an MMF approach, our proposal
shares some of the principles of similar methods for 2D videos.
However, it:
i)~takes into account the specific features of 360-degree videos;
ii)~uses a different set of individual spatial and temporal features;
iii)~is based on an improved temporal pooling method; and
iii)~uses a random forest regression model.
Considered together, those features allows our method to
support a better correlation to subjective scores and more robust results than
state-of-the-art method.

\section{Viewports-based MMF for 360 VQA}
\label{sec:proposed_method}

Fig.~\ref{fig:proposed_method} shows our proposed 360-VQA approach.
The possible space of visible viewports is represented by using $N$ viewports
rendered from different viewing directions.
The same $N$ viewports are rendered from both original and the distorted video
content, and 2D objective metrics are computed individually within the
viewports and then temporally pooled using an HVS-based method.
Finally, based on the per-viewport pooled scores, we train a regression model that
is able to learn a combination of the individual objective metrics into a new
objective metric that closely relates to subjective scores.

The rest of this section details each of the steps above.
In what follows, let $\mathcal{R}=\{R_f, f=0,1,...,F-1\}$ and
$\mathcal{D}=\{D_f,f=0,1,...,F-1\}$ respectively be the reference and
distorted sequences of the same 360-degree video content in the projection
domain.
$R_f$ and $D_f$ denote the $f$'th frame of $\mathcal{R}$ and $\mathcal{D}$,
respectively, and $F$ the total number of frames in both $\mathcal{R}$ and
$\mathcal{D}$.


\subsection{Viewport sampling and field of view}
\begin{figure}[!htb]
\centering
\includegraphics[width=.7\columnwidth]{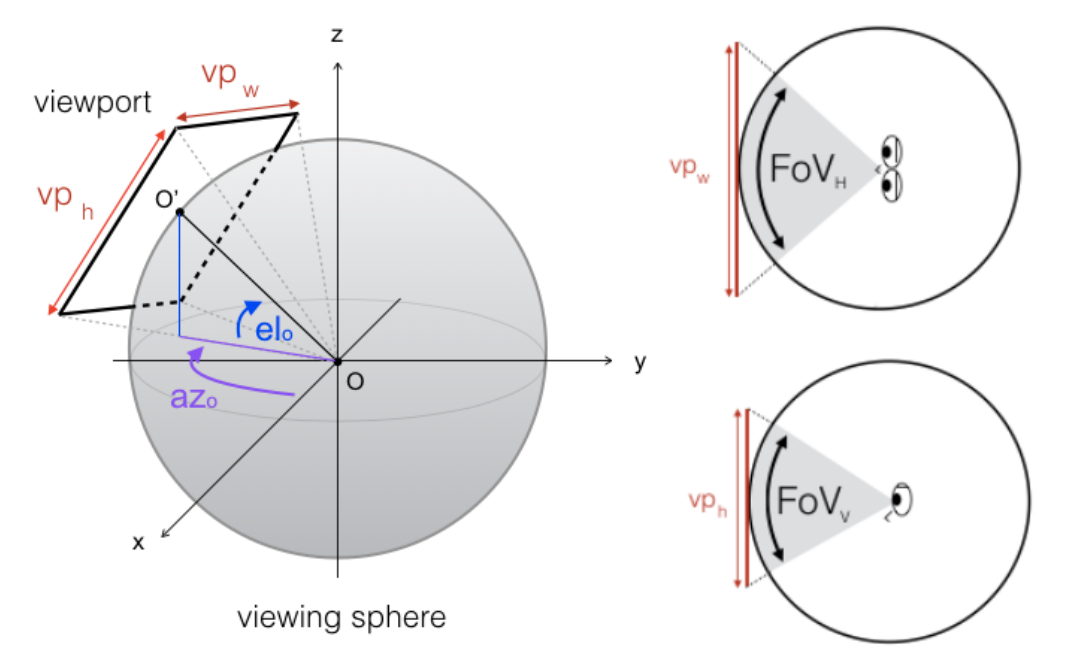}
\caption{Viewport parameters~\cite{desimone_geometry-driven_2016}.}
\label{fig:vp_definition}
\end{figure}

First, for each frame $f$, we compute a set of viewports
${\mathcal{V}}_f^\mathcal{R} =
  \{V_f^{\mathcal{R},0}, ..., V_{f}^{\mathcal{R},N-1}\}$ and 
${\mathcal{V}}_f^\mathcal{D} = 
  \{\mathcal{V}_f^{\mathcal{D},0}, ..., V_{f}^{\mathcal{D}, N-1}\}$,
for the respective reference and distorted frames.
A viewport~(Fig.~\ref{fig:vp_definition}) is the gnomonic
projection~\cite{pearson1990map} of the omnidirectional signal to a plane
tangent to the sphere, which is defined by:
\begin{itemize}
\item the viewing direction~$(el_o, az_o)$, which defines the center $O'$
      where the viewport is tangent to the sphere;
\item its resolution $[vp_w, vp_h]$; and
\item its horizontal and vertical field-of-view, $\text{FoV}_h$ and
      $\text{FoV}_v$, respectively.
\end{itemize}

When considering a viewport-based metric for 360-degree videos, we need to
define a viewport sampling process that, given an omnidirectional image, $I$,
and the viewports parameters $vp_w$, $vp_h$, $FoV_w$ and $FoV_h$, generates
$N$ viewports from different viewing directions~(i.e., different $O'$s).
On one hand, larger FoVs result in both overlapped regions between the
viewports and larger geometry distortions.
Having duplicated content can be an issue because it increases the importance
of such duplicated areas when computing objective metrics on the viewports.
On the other hand, smaller FoVs might require more viewports to completely
cover the sphere area, which might also result in higher computational costs.
Thus, a good balance between the sampling and number of viewports and its
distribution to provide good coverage of the sphere is necessary.
Ideally, the viewport resolution $[vp_w, vp_h]$ should also match the HMD
resolution used to visualize the content.

\begin{figure}[!h]
  \centering
  \includegraphics[width=.80\columnwidth]{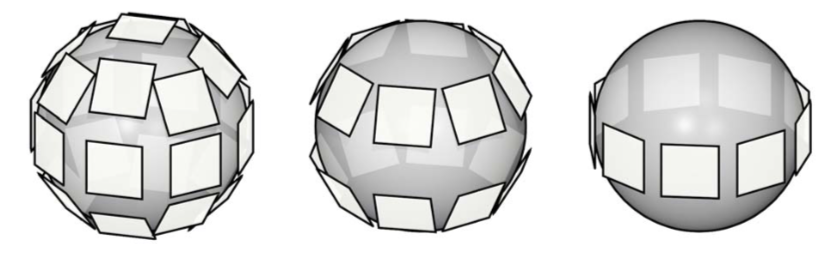}
  \caption{Uniform (left), tropical (center), and equatorial (right)viewports
           sampling for computing viewport-based objective metrics.}
  \label{fig:viewports_schematics}
\end{figure}

\begin{figure}[!htb]
  \centering
  \begin{subfigure}[H]{0.3\textwidth}
    \includegraphics[width=\textwidth]{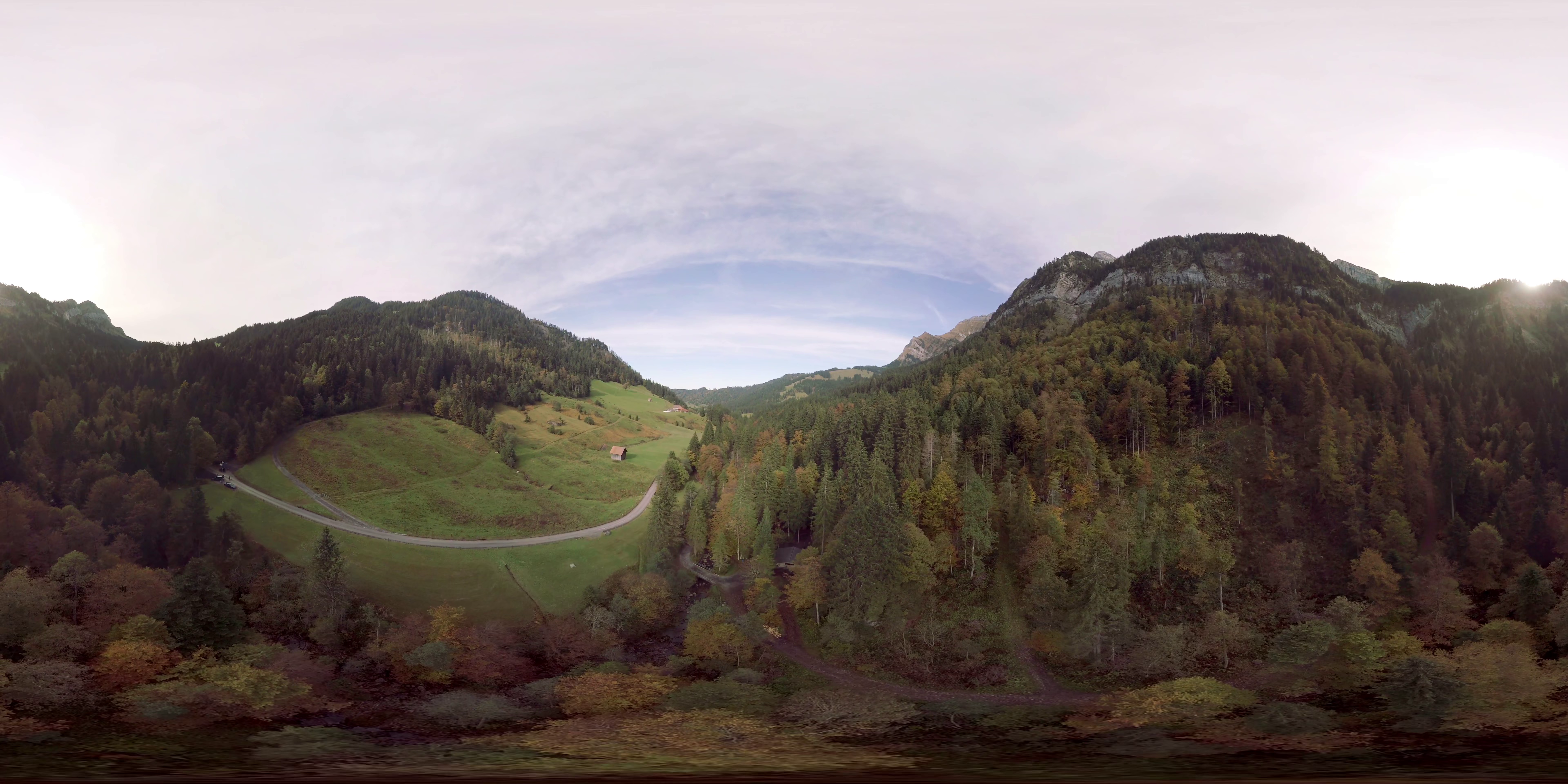}
    \caption{}
    \begin{minipage}{.1cm}
    \vfill
    \end{minipage}
  \end{subfigure}
  ~
  \begin{subfigure}[H]{0.3\textwidth}
    \includegraphics[width=\textwidth]{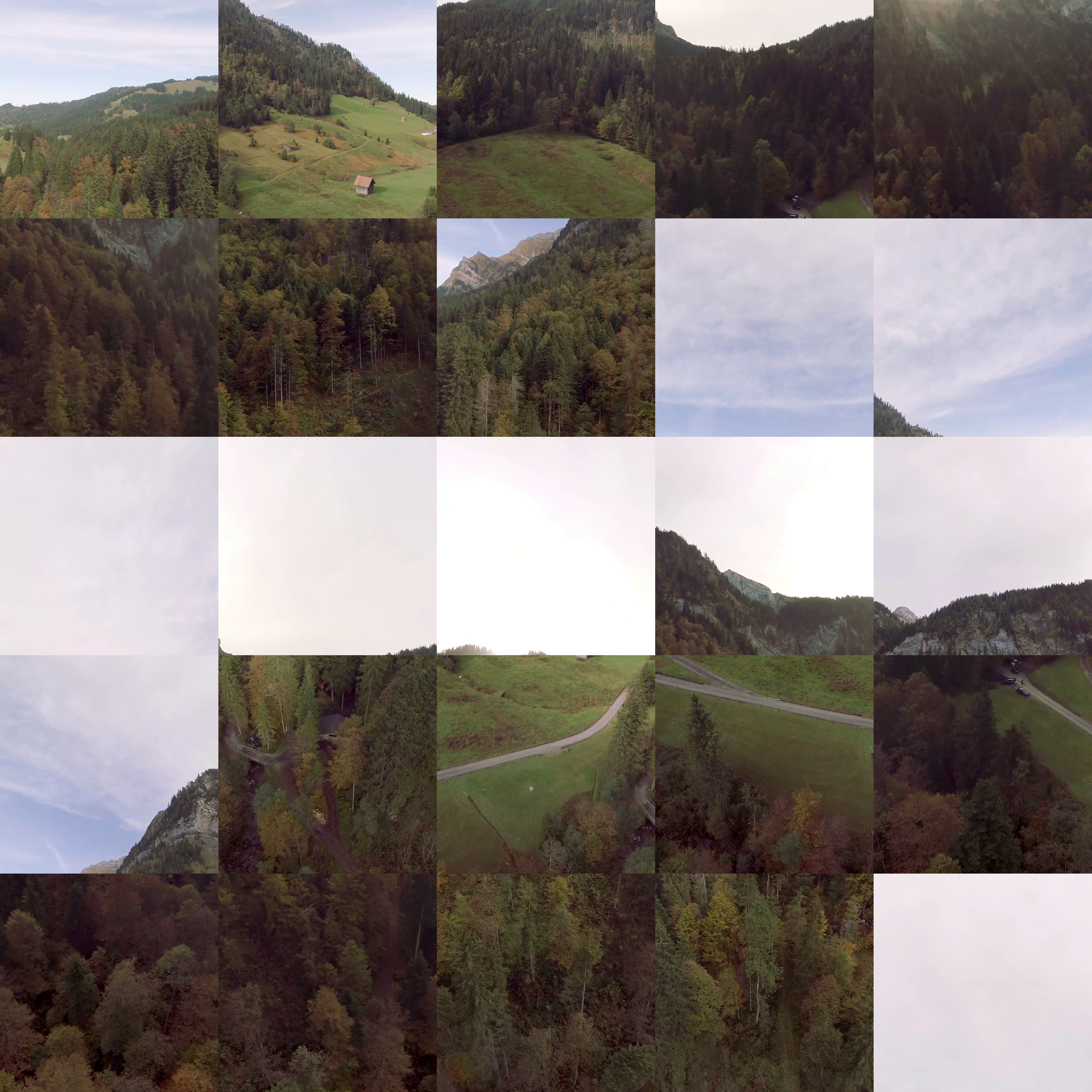}
    \caption{}
  \end{subfigure}
  \caption{Examples of a frame on the ERP projection domain~(a) and an
           aggregate frame with viewports~(b) computed from it using the
           uniform sampling process.}
  \label{fig:panorama_vps_sample}
\end{figure}

Fig.~\ref{fig:viewports_schematics} shows three viewport sampling
configurations: \emph{uniform}, \emph{tropical}, and
\emph{equatorial}~\cite{birkbeck_quantitative_2017}, that sample, respectively,
$25$, $16$, and $9$ viewports.
Fig.~\ref{fig:panorama_vps_sample} shows an example of an ERP frame and the
viewports generated using the \emph{uniform} sampling method.
In the figure, the viewports are aggregated into a single frame that we will
refer to as the \emph{viewports collage} (VP-Collage) frame in the rest of the
paper.

\subsection{Spatial and temporal features}
Based on the previously generated viewports, we compute for each pair of
reference and distorted viewports, $V_f^{\mathcal{R},n}$ and
$V_f^{\mathcal{D},n}$, $0 \leq n < N$, $0 \leq f < F$ a set of $M$ objective
metrics, denoted as:
\begin{equation}
\textbf{Q}_f^n = \{Q_0(V_f^{\mathcal{R},n}, V_f^{\mathcal{D},n}), ..., Q_{m-1}(V_f^{\mathcal{R},n}, V_f^{\mathcal{D},n})\}
\end{equation}
In particular, the following spatial quality metrics are computed for each
viewport pair: 
\begin{itemize}
\item Spatial Activity~(SA),
\item PSNR-HVS and PSNR-HVS-M~\cite{ponomarenko_between_2007},
\item Multi-Scale Structural Similarity~(MS-SSIM)~\cite{wang_multiscale_2003},
      and
\item Gradient Magnitude Similarity Deviation~(GMSD)~\cite{xue_gradient_2014}
\end{itemize}
and the following temporal quality metrics:
\begin{itemize}
\item Relative change in the temporal information~(R-TI), and
\item Temporal Gradient Magnitude Similarity Deviation~(T-GMSD)
\end{itemize}
The above metrics were selected mainly because it has been shown
that they correlated well with subjective scores for traditional image quality
assessment and they complement each other with regards to different 2D image
distortions~\cite{xue_gradient_2014}.
Nevertheless, the proposed method can also be easily adapted to other metrics,
which, for instance, better reflect the distortions in a specific dataset.
The rest of this subsection details each of the above metrics.

\subsubsection{Spatial Activity}
The spatial activity~(SA) of a pair of frames is defined as the
root mean square~(RMS) difference between the Sobel maps of each of the
frames~\cite{freitas_using_2018}.
The Sobel operator, $S$, is defined as:

\begin{equation}
S(z) = \sqrt{(G_1 * z)^2 + (G_1^T * z)^2}
\end{equation}
where $z$ is the frame picture and $*$ denotes the 2-dimensional convolution
operation, $G_1$ is the vertical Sobel filter, given by:
\begin{equation}
G_1 =
\begin{bmatrix}
1 & 0 & -1 \\
2 & 0 & -2 \\
1 & 0 & -1 \\
\end{bmatrix}
\end{equation}
and $G_1^T$ is the transpose of $G_1$ (horizontal Sobel filter).

Let $u$ and $v$ be $V^{\mathcal{R},n}_f$ and $V^{\mathcal{D},n}_f$, i.e., the
same viewport sampled from the same frame $f$ from both reference and distorted
content, respectively.
We define the difference between the Sobel maps of both frames as:
\begin{equation}
s = S(u) - S(v).
\end{equation}
Then, we compute SA as:
\begin{equation}
SA(v,u) = \sqrt{\frac{1}{M N} \sum_{i,j}{|s_{ij}|^2}}
\end{equation}
where $i$, $j$ are respectively the horizontal and vertical indices of $s$,
and $M$ and $N$ are the height and width of the viewports, respectively.

\subsubsection{PSNR-HVS and PSNR-HVS-M}
PSNR-HVS~\cite{egiazarian2006new} and
PSNR-HVS-M~\cite{ponomarenko_between_2007} are two models that have been
designed to improve the performance of PSNR taking into consideration the HVS
properties.
PSNR-HVS divides the image into 8x8 pixels non-overlapping blocks.
Then the $\delta(i,j)$ difference between the original and the distorted
blocks is weighted for every 8x8 block by the coefficients of
the Contrast Sensitivity Function~(CSF).
PSNR-HVS-M~\cite{ponomarenko_between_2007} is defined similarly, but the
difference between the DCT coefficients is further multiplied by a contrast
masking metric~(CM) for every 8x8 block.

\subsubsection{MS-SSIM}
The structural similarity (SSIM) metric divides the job of computing the
similarity between two images into three comparisons: luminance, contrast, and
structure, respectively defined as:

\begin{equation}
l(\mathbf{x},\mathbf{y}) = \frac{2\mu_x\mu_y + C_1}{\mu_x^2 + \mu_y^2 + C_1}
\end{equation}

\begin{equation}
c(\mathbf{x},\mathbf{y}) = \frac{2\sigma_x\sigma_y + C_2}{\sigma_x^2 + \sigma_y^2 + C_2}
\end{equation}

\begin{equation}
s(x,y) = \frac{\sigma_{xy} + C_3}{\sigma_x\sigma_y + C_3}
\end{equation}
where $\mu_x$ and $\mu_y$ denote the mean luminance intensities of the
compared image signals $\mathbf{x}$ and $\mathbf{y}$;
$\sigma_x$ and $\sigma_y$ are the standard deviations of the luminance samples
of the two images;
$\sigma_{xy}$ is the covariance of the luminance samples;
and $C_1$ and $C_2$ are stabilizing constants.
For an image with a dynamic range $L$, $C_1 = (K_1L)^2$ where $K_1$ is a small
constant such that $C_1$ only takes effect when $(\mu_x^2+\mu_y^2)$ is small.
The SSIM index is then defined as:
\begin{equation}
SSIM(\mathbf{x},\mathbf{y}) = [l(\mathbf{x},\mathbf{y})]^{\alpha} \cdot
                              [c(\mathbf{x},\mathbf{y})]^{\beta} \cdot
                              s(\mathbf{x},\mathbf{y})]^{\gamma}
\end{equation}
where $\alpha$, $\beta$, and $\gamma$ are positive parameters that adjust the
relative importance of the three comparison functions.
Setting $\alpha = \beta = \gamma = 1$ and $C_3 = C_2/2$ gives the specific
form:
\begin{equation}
SSIM(x,y) = \frac{(s\mu_x\mu_y + C_1)(2\sigma_{xy} + C2)}
                 {(\mu^2 + \mu_y^2 + C1)(\sigma_x^2 + \sigma_y^2 + C2)}
\end{equation}

MultiScale-SSIM (MS-SSIM) is an extension of SSIM for multiple scales.
At every scale, MS-SSIM applies a low pass filter to the reference and
distorted images and downsample the filtered images by a factor of two.
At the $m$th scale, contrast and structure terms are taken into account:

\begin{equation}
MSSSIM(\mathbf{x},\mathbf{y}) = [l_M(\mathbf{x},\mathbf{y})]^\alpha \cdot
      \prod_{m=1}^M{[c_m(\mathbf{x},\mathbf{y})]^\beta \cdot
[s_m(\mathbf{x},\mathbf{y})]^\gamma}
\end{equation}

\subsubsection{GMSD}
Gradient Magnitude Similarity Deviation~(GMSD)~\cite{xue_gradient_2014} is
based on the standard deviation of the gradient magnitude similarity map, GMS,
which is computed as:

\begin{equation}
GMS(u,v) = \frac{2 \cdot m(u) \cdot m(v) + c}{m(u)^2 + m(v)^2 + c}
\end{equation}
where $u$ and $v$ are respectively the current and previous frame; $c$ is a
positive constant that guarantees stability; and $m(z)$ is:

\begin{equation}
m(z) = \sqrt{(z * G_2)^2 + (z * G_2^T)^2}
\end{equation}
where $*$ denotes the convolution operator, $G_2$ represents the vertical
Prewitt filter:

\begin{equation}
G_2 = \begin{bmatrix}
\frac{1}{3} & 0 & -\frac{1}{3} \\
\frac{1}{3} & 0 & -\frac{1}{3} \\
\frac{1}{3} & 0 & -\frac{1}{3} \\
\end{bmatrix}.
\end{equation}
$G_2^T$ is the transpose of $G_2$, i.e, the horizontal Prewitt filter.
The GMSD index is then computed as:
\begin{equation}
GMSD(u,v) = \sqrt{\frac{1}{NM}\sum_{i,j}(GMS(u, v) - \overline{GMS(u,v)})^2},
\end{equation}
where $\overline{GMS(u,v)}$ is the gradient magnitude similarity mean, computed
as:

\begin{equation}
\overline{GMS(u,v) }= \frac{1}{NM}\sum_{i,j}{GMS(u,v)}
\end{equation}


\subsubsection{Relative change in temporal information}
Temporal information~(TI)\cite{itu_p910} is an indicator that characterizes the
amount of motion in a video and is defined as the standard deviation of the
difference between two frames:

\begin{equation}
\Delta F_n = F_n - F_{n-1}
\end{equation}

\begin{equation}
{TI}[F_n] = std(\Delta F_n)
\end{equation}

Here, we define the relative change in the temporal information as:

\begin{equation}
{TI}_{rel}[F_n] = \frac{|{TI}_{ref}[F_n] - {TI}_{dist}[F_n]|}{{TI}_{ref}[F_n]},
\end{equation}
where ${TI}_{ref}[F_n]$ and ${TI}_{dist}[F_n]$ are respectively the TI for the
frame $F_n$ in the reference and distorted videos.

%
%
%

\subsubsection{Temporal GMSD} is defined as the GMSD score between the
difference of two consecutive references frames and two consecutive distorted
frames, i.e.:

\begin{equation}
  \Delta F_{r}(f) = F_{r}(f) - F_{r}(f-1)
\end{equation}

\begin{equation}
  \Delta F_{d}(f) = F_{d}(f) - F_{d}(f-1)
\end{equation}



\begin{equation}
  \text{T-GMSD}(f) = GMSD(\Delta F_r(f), \Delta F_d(f))
\end{equation}

\subsection{Temporal feature pooling}

The per-viewports metrics $\textbf{Q}_f^n$ for each frame, $f \in \{1, ...,
F\}$ and viewport $n \in \{1, ..., N\}$, are integrated to yield the overall
quality of each viewport: $\textbf{Q}_{pool}^n$.
%
%
This integration is performed by the temporal pooling module.
Our proposal is modular and can be adapted to different temporal pooling
methods.
Based on the experiments of Section~\ref{sec:temporal_pooling_analysis}, and
inspired by~\cite{lu_low-complexity_2019}, we propose a temporal pooling method
considering the characteristics of the HVS, in particular:
\begin{itemize}
\item \emph{the smooth effect}, i.e., the subjective ratings of the whole video
      sequence typically demonstrate far less variations than the frame-level
      quality scores;

\item \emph{the asymmetric effect}, i.e., HVS is more sensitive to frame-level
      quality degradation than to improvement; and

\item \emph{recency effect}, i.e, subjects tend to put a higher weight on what
      they have seen most recently.
\end{itemize}

\begin{figure}[ht!]
  \centering
  \includegraphics[width=0.8\columnwidth]{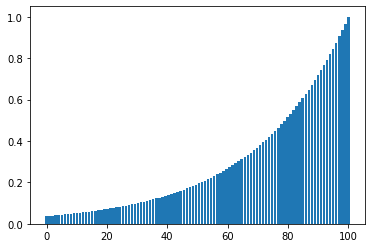}
  \caption{Exponential weights used to average per-frame quality metrics.}
  \label{fig:temporal_pooling_exp_weights}
\end{figure}

More precisely, for each viewport $n$, we first process the original scores
considering both smooth and asymmetric effects as:

\begin{equation}
Q_{LP}^n(f) =
\begin{cases}
Q_{LP}^n(f-1) + \alpha \cdot \Delta Q(f), \text{if } \Delta Q^n \leq 0\\
Q_{LP}^n(f-1) + \beta \cdot \Delta Q(f), \text{if } \Delta Q^n > 0\\
\end{cases}
\end{equation},
where $\Delta Q^n = Q_{frame}^n(f) - Q_{LP}^n(f-1)$ and $Q_{LP}^n =
Q_{frame}^n(1)$, and  $\alpha$ and $\beta$ control the asymmetric weights.
Then, we perform a weighted-average sum of the above processed scores
considering the recency effect:

\begin{equation}
Q_{pool}^n = \frac{1}{F}\sum_{f=1}^{F}{Q^n_{LP}(f) w(f)}.
\end{equation}
Here, we define the weights $w(f)$ as an exponential function~(see
Fig.~\ref{fig:temporal_pooling_exp_weights}):

\begin{equation}
w(f) = e^{(((f+1)-F)/\tau)}.
\end{equation}

Similar to~\cite{lu_low-complexity_2019}, in our experiments
we use $\alpha=0.03$ and $\beta=0.2$.

\subsection{Regression}

After the temporal pooling, we end up with $M$ features for each viewport,
which are then concatenated as a feature vector,
$
\textbf{Q} = [Q_0^0, ..., Q_{m-1}^0, Q_0^1, ..., Q_{m-1}^1, Q_0^{n-1}, ..., Q_{m-1}^{n-1}]
$.
Such a vector is used for learning a non-linear mapping between the computed
per-viewport features and the subjective DMOS scores of 360-degree videos.
In our framework, we have tested three different regression methods:
\begin{itemize}
  \item Support Vector Regression~(SVR)~\cite{basak2007support};
  \item Gradient Boosting regression~(GBR)~\cite{friedman2001greedy}; and
  \item Random Forest Regression~(RFR)~\cite{breiman2001random}.
\end{itemize} 

Based on the experiments on Section~\ref{sec:experimental_setup}, we have
chosen RFR as our final regression method because it significantly outperformed
the other methods.
Next we detail our experiments setup, hyper-parameter tuning, training, and
test processes.

\section{Experimental results and analysis}
\label{sec:experimental_setup}

Unless specified otherwise, we validate  our proposal based on the
VQA-ODV~\cite{li_bridge_2018} dataset.
VQA-ODV is the largest publicly available dataset today and is composed of
three types of projections, ERP, RCMP, and TSP, and 3-levels of H.265
distortions, quantization parameters~(QP)=27, 37, and 42.
In total, there are 60 different reference sequences (12 in raw format and
others downloaded from YouTube VR channel) and 540 distorted sequences that
were rated by 221 participants.
In all our following experiments, we extract only the ERP sequences from
VQA-ODV, resulting in 180 distorted sequences.
Both MOS~(Mean Opinion Scores) and DMOS~(Differential Mean Opinion Scores) are
available for the dataset.


We compare our method to PSNR, S-PSNR, WS-PSNR MS-SSIM, and VMAF, using common
criteria for the evaluation of objective quality metrics:
Pearson Linear Correlation Coefficient~(PLCC),
Spearman Rank Order Correlation Coefficient~(SROCC),
and Root Mean Squared Error~(RMSE).
SROCC measures the prediction monotonicity while PLCC and RMSE measure the
prediction accuracy.
Higher SROCC, PLCC and lower RMSE indicate good correlation with subjective
scores.

Moreover,
we compare the performance of our method and the other objective
metrics when the features are computed in:
i)~the projection domain~(``Proj.'');
ii)~all viewports merged in a collage frame (``VP-Collage'')~(see
Fig.~\ref{fig:panorama_vps_sample}); and
iii)~the viewports considered indiviually ("VP"), i.e., the metrics are
computed independently for each viewports~(as discussed in
Section~\ref{sec:proposed_method}).
Computing the objective metrics in the viewport collage frame is similar to
averaging the quality of all the viewports.

Based on the above 3 different modes, we performed the following experiments:
\begin{itemize}
  \item using the same fixed train/test subset of the VQA-ODV dataset used
        in~\cite{li_bridge_2018,li_viewport_2019}~(Subsection~\ref{sec:fixed_train_test_experiment};
  \item a cross-validation experiment in the whole ERP sequences of VQA-ODV
        dataset~(Subsection~\ref{sec:cross_validation_experiment}); and
  \item a cross dataset validation~\ref{sec:cross_dataset_experiment}.
\end{itemize}

In all the above cases, we perform grouped split of the data.
Such approach divides the database into two content-independent subsets
(training and testing) ensuring that videos generated from one reference (i.e.,
the same content) in the testing subset are not present in the training subset,
and vice-versa.
Also, in the above following, we use a uniform sampling with a 40-deg field
of view for the viewports, which resolution matches the HMD resolution used in
the dataset~(an HTC Vive).

\color{black}

\subsection{VQA-ODV: Fixed train/test subsets}
\label{sec:fixed_train_test_experiment}

Table~\ref{tbl:fixed_train_test_results} shows the results of our method using
the same~(fixed) train/test selection
of~\cite{li_bridge_2018,li_viewport_2019}, which ensures that the same
sequence content is not part of both train and test sets.
After separating the dataset into train and test sets, we first run a group
shuffle cross-validation on the training data to find the best random forest
hyper-parameters to predict the 360-VQA.
Based on the resulting hyper-parameters we then train and validate the model,
respectively using the previously separated test data.
Then, we compute the performance of the model to predict the scores of the
sequences in the test set.
By following the above procedure, we make sure that the
hyper-parameter tunning never sees the test data.

We compare the performance of our method against: PSNR, MS-SSIM, and
VMAF~(state-of-the-art metrics for 2D quality assessment) when the features
are computed in both the projection and the viewports domain; and S-PSNR and
WS-PSNR, metrics specifically developed for 360 VQA.
For the non-learned metrics~(PSNR, MS-SSIM, S-PSNR, and WS-PSNR), we emulate
the training phase by fitting the following 4-parameter logistic function with
the train set and then computing its performance with the test set:

\begin{equation}\label{eq:logit4}
s^{\prime} = \frac{\beta_1 - \beta_2}{1 + e^{-\frac{S-\beta_3}{||\beta_4||}}}
+ \beta_2.
\end{equation}

Finally, for our method, we compare different regression
techniques: Support Vector Regression~(SVR), Gradient Boosting
Regression~(GBR), and Random Forest Regression~(RFR).

\textbf{Discussion.} From
Table~\ref{tbl:fixed_train_test_results}, the best performance for the fixed
train/test set on VQA-ODV is achieved by our method using features computed
separatly for each viewport (``VP'' mode).
Such results can be explained by both the viewports being closer to what the
users see and the model being able to learn the most important viewports, which
is not the case when using the ``VP-Collage'' mode.
Our results in Table~\ref{tbl:fixed_train_test_results} also show that
computing objective metrics on the viewport domain~(``VP-Collage'') improve the
performance of all the tested metrics~(LCC gain around $0.05$) validating our
hypothesis that viewports better represents the perceived quality when users
watch a 360 video on an HMD when compared to the projection domain.
Finally, it is also interesting to note that our method (on both ``Proj.'' and
``VP-Collage'') outperforms VMAF, which can be explained by the choice of
objective metrics, the temporal pooling, and regression methods in our method.
Finally, with regards to generalization of our method,
Table~\ref{tbl:train_fixed_train_test_set_performance} also brings the training
performance of our different regression methods.
When comparing Table~\ref{tbl:fixed_train_test_results}
and~\ref{tbl:train_fixed_train_test_set_performance} it is clear that RFR is
the most robust regression method among the tested ones, providing PLCC values
while avoiding overfitting.

\begin{table}[!htb]
\centering
\caption{Fixed train/test set test results (VQA-ODV dataset).
}
\label{tbl:fixed_train_test_results}
\resizebox{.95\columnwidth}{!}{
\begin{tabular}{l|n{2}{3}n{2}{3}n{2}{3}}
\toprule
Metric & {PLCC~$\uparrow$} & {SROCC~$\uparrow$} & {RMSE~$\downarrow$}
\\ \midrule
PSNR
& 0.72495
& 0.73797
& 8.1760
\\
PSNR~(VP-Collage)
& 0.76222
& 0.76345
& 7.5824
\\
S-PSNR
& 0.75138
& 0.77040
& 7.7557
\\
WS-PSNR
& 0.74328
& 0.56056
& 7.9501
\\
MS-SSIM~(Proj.)
& 0.76005
& 0.78867
& 7.8741
\\
MS-SSIM~(VP-Collage)
& 0.81719
& 0.84144
& 7.0024
\\
VMAF~(Proj.)
& 0.79657
& 0.79382
& 7.2481
\\
VMAF~(VP-Collage)
& 0.84483
& 0.85637
& 6.2710
\\
\midrule
Ours~(SVR, Proj.)
& 0.80422
& 0.83552
& 7.2508
\\
Ours~(SVR, VP-Collage)
& 0.85821
& 0.9045
& 8.1205
\\
Ours~(SVR, VP) (Overfitting)
& 0.51266
& 0.72432
& 10.412
\\ \midrule

Ours~(GB, Proj.)
& 0.8467
& 0.84994
& 6.474
\\
Ours~(GB, VP-Collage)
& 0.85727
& 0.85354
& 6.2516
\\
Ours~(GB, VP)
& 0.91462
& 0.88906
& 5.0407
\\ \midrule
Ours~(RFR, Proj.)
& {\npboldmath} 0.85629
& {\npboldmath} 0.86873
& {\npboldmath} 6.3588
\\
Ours~(RFR, VP-Collage)
& {\npboldmath} 0.90559
& {\npboldmath} 0.89318
& {\npboldmath} 5.55
\\
Ours~(RFR, VP)
& {\npboldmath} 0.9293
& {\npboldmath} 0.91663
& {\npboldmath} 4.5614
\\
\bottomrule
\end{tabular}}
\end{table}

\begin{table}[!hbt]
\center
\caption{Training performance of the different regression methods
         on the fixed train set of VQA-ODV.} 
\label{tbl:train_fixed_train_test_set_performance}
\resizebox{0.95\columnwidth}{!}{
\begin{tabular}{l | n{2}{3} n{2}{3} n{2}{3}}
\toprule
Metric & {PLCC~$\uparrow$} & {SROCC~$\uparrow$} & {RMSE~$\downarrow$} \\ \midrule
Ours~(SVR, Proj.).      & 0.75075 & 0.77536 & 7.7877 \\
Ours~(SVR, VP-Collage)  & 0.78099 & 0.80983 & 7.4548 \\
Ours~(SVR, VP)          & 0.99999 & 0.99997 & 0.080053 \\
\midrule
Ours~(GB, Proj.).     & 0.99025 & 0.98925 & 1.7368 \\
Ours~(GB, VP-Collage) & 0.98936 & 0.9873 & 1.7979 \\
Ours~(GB, VP)         & 0.99946 & 0.9987 & 0.41038 \\
\midrule
Ours~(RFR, Proj.).     & 0.93477 & 0.94597 & 4.2587 \\
Ours~(RFR, VP-Collage) & 0.96616 & 0.97202 & 3.2244 \\
Ours~(RFR, VP)         & 0.98256 & 0.98399 & 2.42 \\
\bottomrule
\end{tabular}}
\end{table}

\subsection{VQA-ODV: Cross-validation}
\label{sec:cross_validation_experiment}

\begin{figure}[!hbt]
\centering
\includegraphics[width=\columnwidth]{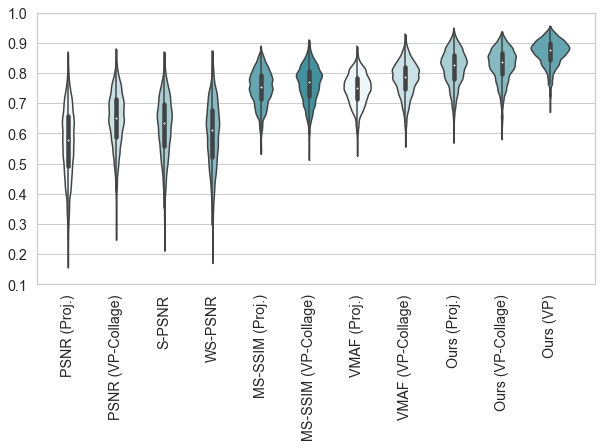}
\caption{Violin plots for the GroupShuffle (80\%/20\%) cross-validation PLCC performance
         on VQA-ODV dataset.}
\label{fig:plcc_distribution}
\end{figure}

To avoid bias on the specific train/test set used above, we also performed a
full cross-validation on the VQA-ODV dataset.
In the cross-validation experiments, we performed a 1000x randomly group
selection of 80\%/20\% train/test splitting of the dataset, and then computed
the average PLCC, SROCC, and RMSE of the models.
The group selection ensures that there is no overlap between content in the
training and test sets.
The hyper-parameters used for the RF regression are the same ones found on the
fixed train/test experiments above.
Table~\ref{tbl:cross_validation_results_80_20} shows the average PLCC, SROCC,
and RMSE results for the cross-validation experiments, and
Fig.~\ref{fig:plcc_distribution} depicts the distribution of the correlation
scores through a violin plot.
Larger sections of the violin plots depict a higher probability of achieving
these correlation scores, while narrower sections depict a lower probability.

\begin{table}[!htb]
\caption{Average of GroupShuffle cross validation~(80\%/20\%) performance on
         VQA-ODV.}
\label{tbl:cross_validation_results_80_20}
\centering
\resizebox{.95\columnwidth}{!}{
\begin{tabular}{l|n{2}{3}n{2}{3}n{2}{3}}
\toprule
Metric & {PLCC~$\uparrow$} & {SROCC~$\uparrow$} & {RMSE~$\downarrow$}
\\ \midrule
PSNR~(Proj.)
& 0.57156
& 0.61873
& 9.8249
\\
PSNR (VP-Collage)
& 0.64746
& 0.68579
& 9.1224
\\
S-PSNR
& 0.62460
& 0.66731
& 9.3461
\\
WS-PSNR
& 0.59803
& 0.64501
& 9.5983
\\
MS-SSIM (Proj.)
& 0.75004
& 0.77535
& 7.9351
\\
MS-SSIM (VP-Collage)
& 0.76405
& 0.79113
& 7.758
\\
VMAF
& 0.74692
& 0.76673
& 7.9631
\\
VMAF (VP-Collage)
& 0.78085
& 0.79802
& 7.5147
\\ \midrule
Ours~(RFR, Proj.)
& {\npboldmath}0.81728
& {\npboldmath}.82901
& {\npboldmath}6.8716
\\
Ours~(RFR, VP-Collage)
& {\npboldmath}0.82676
& {\npboldmath}0.82647
& {\npboldmath}6.7376
\\
Ours~(RFR, VP)
& {\npboldmath}0.86778
& {\npboldmath}0.86769
& {\npboldmath}5.9367
\\ \bottomrule
\end{tabular}
}
\end{table}

\textbf{Discussion.}
Table~\ref{tbl:cross_validation_results_80_20} further validates our previously
conclusions from the fixed train/test dataset settings, namely:
i)~considering viewports~(``VP-Collage'' and ``VP'' modes) instead of the
projection domain improves the performance of objective quality metrics;
ii)~considering features on individual viewpors~(``VP'' mode) allows the model
to further weight the different views based on the importance of those regions
to the final quality of the 360 video.
Moreover, Figure~\ref{fig:plcc_distribution} shows that besides having a better
average, our method also provide a higher density on the high PLCC values.

\subsection{VQA-ODV x VR-VQA48: Cross dataset validation}
\label{sec:cross_dataset_experiment}

To demonstrate the perfomance of our proposal in more than one dataset,
Table~\ref{tbl:cross_dataset_performance} shows
the results of the performance of our model trained on the ERP sequences of
VQA-ODV and tested on public available VR-VQA48 dataset~\cite{xu_assessing_2018}.
VR-VQA48 is composed of $12$ original omnidirectional video sequences (YUV
4:2:0 format at the resolution of $4096\times2048$) and $36$ corresponding impaired
sequences by encoding each original sequence with $3$ different bitrate
settings.
48 subjects rated raw subjective quality scores for all the 48 sequences.
MOS and DMOS values are available for the sequences per subject.
We consider the final quality score of each sequence as the average DMOS.
As in the previous setups, we also use the uniform viewport sampling with a
40-degree of field-of-view.
In total, the model is trained on the $180$ ERP sequences of and tested on the
$36$ distorted sequences of VR-VQA48.
Also, from the previous tables, Table~\ref{tbl:cross_dataset_performance} also
add the results of the method proposed by~\cite{gao2020quality} as reported in
the original work.

\begin{table}[!hbt]
\center
\caption{Objective metrics performance on VR-VQA48. Our method is
         trained on VQA-ODV and tested on VR-VQ48}
\label{tbl:cross_dataset_performance}
\resizebox{0.95\columnwidth}{!}{
\begin{tabular}{l | n{2}{3} n{2}{3} n{2}{3}}
\toprule
Metric & {PLCC~$\uparrow$} & {SROCC~$\uparrow$} & {RMSE~$\downarrow$} \\ \midrule
PSNR~(Proj.)         & 0.499 & 0.508 & 10.732 \\
PSNR~(VP-Collage)    & 0.613 & 0.612 & 9.843  \\
S-PSNR               & 0.569 & 0.595 & 10.183 \\
CPP-PSNR             & 0.567 & 0.595 & 10.198 \\
WS-PSNR              & 0.548 & 0.562 & 8.804  \\
MS-SSIM~(Proj.)      & 0.75864 & 0.75067 & 8.4901 \\
MS-SSIM~(VP-Collage) & 0.843 & 0.829 & 7.634  \\
VMAF~(Proj.)         & 0.783 & 0.771 & 7.712  \\

\midrule
OV-PSNR[PSNR]~\cite{gao2020quality}     & 0.837 & 0.890 & 6.749 \\
OV-PSNR[S-PSNR]~\cite{gao2020quality}   & 0.818 & 0.775 & 7.123 \\
OV-PSNR[CPP-PSNR]~\cite{gao2020quality} & 0.837 & 0.787 & 5.181 \\
OV-PSNR[WS-PSNR]~\cite{gao2020quality}  & 0.838 & 0.790 & 5.157 \\

\midrule
Ours~(SVR, Proj.).     & 0.79504 & 0.77503 & 7.8983 \\
Ours~(SVR, VP-Collage) & 0.87599 & 0.85483 & 6.9807 \\
Ours~(SVR, VP)~(overfitting) & 0.69851 & 0.77497 & 9.5589 \\

\midrule
Ours~(GB, Proj.).     & 0.78669 & 0.78049 & 8.2379 \\
Ours~(GB, VP-Collage) & 0.89119 & 0.88057 & 6.0305 \\
Ours~(GB, VP)         & 0.94896 & 0.93951 & 5.7236 \\

\midrule
\textbf{Ours~(RFR, Proj.)}      & {\npboldmath}0.83652 & {\npboldmath}0.82033 & {\npboldmath}7.6893 \\
\textbf{Ours~(RFR, VP-Collage)} & {\npboldmath}0.92515 & {\npboldmath}0.89961 & {\npboldmath}5.4293 \\
\textbf{Ours~(RFR, VP)}         & {\npboldmath}0.95644 & {\npboldmath}0.94852 & {\npboldmath}5.1607 \\
\bottomrule
\end{tabular}}
\end{table}

\begin{table}[!hbt]
\center
\caption{Training performance of the different regression methods
trained on VQA-ODV.}
\label{tbl:train_cross_dataset_performance}
\resizebox{0.95\columnwidth}{!}{
\begin{tabular}{l | n{2}{3} n{2}{3} n{2}{3}}
\toprule
Metric & {PLCC~$\uparrow$} & {SROCC~$\uparrow$} & {RMSE~$\downarrow$} \\ \midrule
Ours~(SVR, Proj.).           & 0.77063 & 0.79571 & 7.5231 \\
Ours~(SVR, VP-Collage)       & 0.79159 & 0.81463 & 7.2407 \\
Ours~(SVR, VP) (Overfitting) & 0.99999 & 1.0 & 0.079993 \\
\midrule
Ours~(GB, Proj.).     & 0.98636 & 0.82033 & 7.6893 \\
Ours~(GB, VP-Collage) & 0.9829  & 0.97905 & 2.2395 \\
Ours~(GB, VP)         & 0.99928 & 0.99882 & 0.47042 \\
\midrule
Ours~(RFR, Proj.).     & 0.94494 & 0.82033 & 7.6893 \\
Ours~(RFR, VP-Collage) & 0.97003 & 0.97341 & 3.0017 \\
Ours~(RFR, VP)         & 0.98658 & 0.98665 & 2.1085\\
\bottomrule
\end{tabular}}
\end{table}

\textbf{Discussion.} The results in Table~\ref{tbl:cross_dataset_performance}
confirm our previous conclusions, showing that our method is also robust across
different datasets.
Of course, if the distortions included in a new dataset are not well modeled by
the individual features of our model, we would not expect to achieve good
results on that dataset.
That does not seem the case for the VR-VQA48 dataset.
Table~\ref{tbl:train_cross_dataset_performance} brings the training
performance, further validating that RFR is the best regression method among
the tested ones.

%

\section{Ablation studies}
In this section we provide additional experiments that allow us to better
understand the influence of viewport
sampling~(Subsection~\ref{sec:viewport_sampling_analysis}); the importance of
different individual features~(Subsection~\ref{sec:ablation_studies}); and the
influence of different temporal pooling
methods~(Subsection~\ref{sec:temporal_pooling_analysis}).

\subsection{Features selection}
\label{sec:ablation_studies}

To study the importance of each feature for our model we performed additional
experiments based on the Sequential Forward Feature Selection~(SFFS) method.
In such a method, we start with an empty set of features ($M=m_i|m=1, ...f$),
and for each step we select the next previously still not selected feature
($m^*$) that maximize a specific metric.
We focus on minimizing PLCC in our experiments.
Table~\ref{tbl:ablation_studies} shows the results for 1 to 4 selected features
on the fixed train/test setup.
As can be seen in Table~\ref{tbl:ablation_studies}, we can improve even further
our results with just a subset of the initially proposed features.
By having a subset of features, however, it is possible that the model does not
generalize as well to other distortion types.
More experiments, with datasets including other distortions type, should be
performed in the future.
One of the main issues that prevented us for performing such experiments is the
lack of availability of such datasets, which we see as an important future work
for the research community.

\begin{table}[!htb]
\centering
\caption{Performance of our proposal using SFFS on fixed train/test VQA-ODV
         dataset.}
\label{tbl:ablation_studies}
\resizebox{0.98\columnwidth}{!}{
\begin{tabular}{l|n{2}{3}n{2}{3}n{2}{3}}
\toprule
Selected features~(RFR, Proj.) & {PLCC~$\uparrow$} & {SROCC~$\uparrow$} & {RMSE~$\downarrow$}
\\ \midrule
GMSD
& 0.80779
& 0.77002
& 6.8896
\\
GMSD, R-TI
& 0.87407
& 0.87759
& 5.8311
\\
GMSD, R-TI, PSNR-HVS
& 0.8848
& 0.87928
& 5.726
\\
GMSD, R-TI, PSNR-HVS, PSNR-HVS-M
& 0.89127
& 0.90914
& 5.5066
\\ \midrule

Selected features~(RFR, VP-Collage) & {PLCC~$\uparrow$} & {SROCC~$\uparrow$} & {RMSE~$\downarrow$}
\\ \midrule
PSNR-HVS-M
& 0.84070
& 0.82598
& 6.4697
\\
PSNR-HVS-M, R-TI
& 0.888
& 0.86873
& 5.5855
\\
PSNR-HVS-M, R-TI, GMSD
& 0.91131
& 0.9009
& 5.3716
\\
PSNR-HVS-M, R-TI, GMSD, T-GMSD
& 0.91625
& 0.90064
& 5.3916
\\ \midrule

Selected features~(RFR, VP) & {PLCC~$\uparrow$} & {SROCC~$\uparrow$} & {RMSE~$\downarrow$}
\\ \midrule
PSNR-HVS-M
& 0.87701
& 0.87362
& 5.6951
\\
PSNR-HVS-M, R-TI
& 0.92056
& 0.93359
& 5.2212
\\
PSNR-HVS-M, R-TI, GMSD
& 0.94578
& 0.94337
& 4.9235
\\
PSNR-HVS-M, R-TI, GMSD, T-GMSD
& 0.94558
& 0.93951
& 4.9707
\\ \bottomrule
\end{tabular}
}
\end{table}

\color{black}

\color{black}
\subsection{Influence of viewport sampling}
\label{sec:viewport_sampling_analysis}
\ranote{Or: On the performance of viewport-based metrics for 360-degree video
         quality evaluation}

When considering a viewport-based metric for 360-degree videos there are (in
theory) infinite ways on how to sample the viewports on the sphere.
To better understand the visual quality of objective metrics computed in the
projection domain against the same metrics computed in the viewports, we
report here a study on objective quality metrics on the VQA-ODV dataset.
For that, we computed the following objective metrics in both the projection
and viewport domains:
PSNR, PSNR-HVS, PSNR-HVSM, SSIM, MSSSIM, GMSD, 
Spatial activity, and Temporal activity.
In the viewport domain, we consider different field-of-views and viewports
sampling patterns.
In both cases~(projection and viewport domains) the features are computed
individually for each frame and then pooled with an average method.

For the following experiments, we have chosen the \emph{uniform} and
\emph{tropical} sampling methods shown in
Fig.~\ref{fig:viewports_schematics}~\cite{birkbeck_quantitative_2017} with
FoVs of 30, 40, and 50 degrees.
In total, there are $16$ and $25$ viewports for the tropical and uniform
sampling methos, respectively.
Finally, we compute the Pearson and Spearman Correlation Coefficient between
the DMOS and the fitted 4-parameter logistic regression, given by
Equation~\ref{eq:logit4}.

\begin{figure*}[h!]
 \centering
  \includegraphics[width=0.75\textwidth]{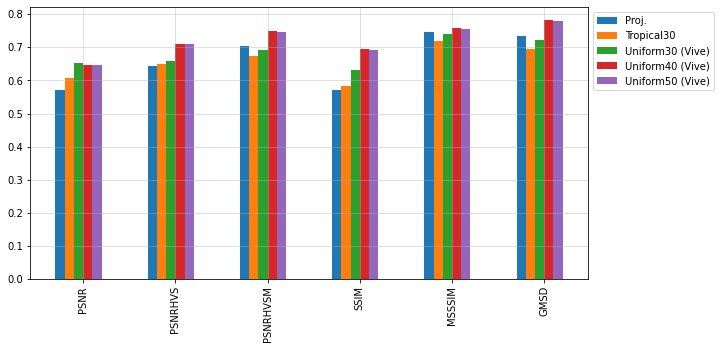}
  \caption{Performance (PLCC) of individual objective viewport metrics. \ranote{Still
           missing Tropical40.}}
  \label{fig:individual_metrics_performance}
\end{figure*}

Table~\ref{tbl:individual_metrics_performance} shows the LCC and SROCC
performance of the different viewport sampling and field-of-views compared
with the same metrics computed in the frame domain.
Fig.~\ref{fig:individual_metrics_performance} plots the LCC values
performance.
From those results, we can conclude that the \emph{uniform} sampling with
40-deg sampling is the one with the best overall performance.

\begin{table*}[h!]
\center
\caption{Individual objective metrics performance on VQA-ODV (mean temporal
         pooling). Tropical and Uniform metrics are computed
	 on VP-Collage frames.}
\label{tbl:individual_metrics_performance}

\resizebox{\textwidth}{!}{
\begin{tabular}{c|n{2}{3}n{2}{3}|n{2}{3}n{2}{3}|n{2}{3}n{2}{3}|n{2}{3}n{2}{3}|n{2}{3}n{2}{3}|n{2}{3}n{2}{3}}
\toprule
\multicolumn{1}{c|}{} &
\multicolumn{2}{|c|}{Proj.~(ERP)} &
\multicolumn{2}{|c|}{Tropical~(30deg)} &
\multicolumn{2}{|c|}{Uniform~(30deg)} &
\multicolumn{2}{|c|}{Tropical~(40deg)} &
\multicolumn{2}{|c|}{Uniform~(40deg)} & 
\multicolumn{2}{|c}{Uniform~(50deg)} 
\\ \midrule

Metric    & LCC~$\uparrow$ & SROCC~$\uparrow$ 
          & LCC~$\uparrow$ & SROCC~$\uparrow$ 
          & LCC~$\uparrow$ & SROCC~$\uparrow$ 
          & LCC~$\uparrow$ & SROCC~$\uparrow$ 
          & LCC~$\uparrow$ & SROCC~$\uparrow$ 
          & LCC~$\uparrow$ & SROCC~$\uparrow$ 
          \\ \midrule

PSNR
& 0.57094 & 0.60200 
& 0.60630 & 0.63945 
& {\npboldmath}0.65222 & {\npboldmath}0.68715 
& 0.62089 & 0.64993 
& 0.64770 & 0.67793 
& 0.64640 & 0.67592 
\\

PSNR-HVS
& 0.64224 & 0.67849 
& 0.64831 & 0.69616 
& 0.65919 & 0.69538 
& 0.63729 & 0.66683 
& {\npboldmath}0.71012 & {\npboldmath}0.74026 
& 0.70981 & 0.74255 
\\

PSNR-HVS-M
& 0.70305 & 0.72821 
& 0.67325 & 0.69414 
& 0.69184 & 0.72514 
& 0.67691 & 0.70345 
& {\npboldmath}0.75087 & {\npboldmath}0.77691 
& 0.74773 & 0.77504 
\\

SSIM      
& 0.57218 & 0.60457 
& 0.58369 & 0.62165 
& 0.63115 & 0.66774 
& 0.59446 & 0.63001 
& {\npboldmath}0.69454 & {\npboldmath}0.72698 
& 0.69124 & 0.72342 
\\

MS-SSIM
& 0.74700 & 0.76860 
& 0.71811 & 0.74878 
& 0.74109 & 0.76924 
& 0.72712 & 0.75607 
& {\npboldmath}0.75733 & {\npboldmath}0.78485 
& 0.74773 & 0.77504 
\\

GMSD
& 0.73405 & 0.75168 
& 0.69441 & 0.71870 
& 0.72169 & 0.74782 
& 0.70259 & 0.72626 
& {\npboldmath}0.78256 & {\npboldmath}0.80594 
& 0.77917 & 0.80109 
\\


VIFP
& {\npboldmath}0.65213 & {\npboldmath}0.71541 
& 0.63339 & 0.67414 
& 0.65386 & 0.71726 
& 0.62946 & 0.63905 
& 0.64378 & 0.67316 
& 0.64373 & 0.67555 
\\


SA
& 0.49929 & 0.53244 
& 0.50804 & 0.54035 
& 0.56426 & 0.59508 
& 0.52184 & 0.55239 
& {\npboldmath}0.66543 & {\npboldmath}0.7008 
& 0.66157 & 0.69406 
\\ \bottomrule



\end{tabular}} 
\end{table*}

\subsection{Influence of temporal pooling}
\label{sec:temporal_pooling_analysis}

Different temporal poolings might also result in different
performance~\cite{Tu2020ACE}.
To better understand the performance of our method when using different
temporal pooling methods, we also performed an experiment on the same
architecture of Figure~\ref{fig:proposed_method} when only changing the
temporal pooling method.
Despite our proposed temporal pooling method, we also tested:

\paragraph{Arithmetic mean}  The sample mean of frame-level scores:
\begin{equation}
Q = \frac{1}{N}\sum^{N}_{n=1}{q_n}.
\end{equation}

\paragraph{Minkowski mean} The $L_p$ Minkowski summation of time-varying
quality is defined as:
\begin{equation}
Q = \left(\frac{1}{N}\sum_{n=1}^{p}{q_n^p}\right)^{1/p}
\end{equation}

\paragraph{Percentile}
Percentile pooling is based on observed phenomenon that perceptual
quality is heavily affected by the ``worst'' parts of the content.
Many prior works have studied and justified (or challenged) percentile pooling
[15– 18, 20].
The k-th percentile pooling is expressed:
\begin{equation}
Q = \frac{1}{|P_{\downarrow k\%}|} \sum_{n \in P_{\downarrow k\%}}{q_n}
\end{equation}

\ranote{Maybe add: Temporal Hysteresis, Narwaria, 2012}

Table~\ref{tbl:temporal_pooling} shows the performance of our method~(VP) using
the different pooling methods on the VQA-ODV dataset.
For the percentile method, we use $k=10$ (i.e., we used only the $10\%$ worst
scores of the frames.
For the Minkowski mean we report the results on both $p=2$ and $p=4$.
From the results, we can conclude that our proposal performs better with our
HVS-based pooling proposed in Section~\ref{sec:proposed_method}.


\begin{table}[!htb]
\centering
\caption{Temporal pooling performance on VQA-ODV fixed train/test sets~(Ours (RFR, VP)).}
\label{tbl:temporal_pooling}
\resizebox{0.85\columnwidth}{!}{
\begin{tabular}{l|n{2}{3}|n{2}{3}|n{2}{3}}
\toprule
Temporal pooling & {PLCC~$\uparrow$} & {SROCC~$\uparrow$} & {RMSE~$\downarrow$}
\\ \midrule
Minkowski4
& 0.90047
& 0.88134
& 5.7978
\\
Minkowski2
& 0.90956
& 0.9027
& 5.3607
\\
Mean
& 0.91828
& 0.90837
& 5.3638
\\ 
Percentile
& 0.9187
& 0.90502
& 5.4154
\\ 
HVS~(Ours)
& {\npboldmath}0.9293
& {\npboldmath}0.9166
& {\npboldmath}4.5614
\\ \bottomrule
\end{tabular}
}
\end{table}

\section{Conclusion}
\label{sec:conclusion}
We propose the use of viewport-based multi-metrics fusion for 360-degree VQA
and discuss the lessons learned by implementing and evaluating such an approach
on two publicly available 360 video datasets.
The computation of features in viewports implies that our metric can be applied
on a variety of projections, and our experiments demonstrate that the MMF
approach is capable of achieving state-of-the-art results while requiring much
less training data than deep learning techniques.

As future work, we plan:
(i)~to consider color and visual attention;
(ii)~to consider fixation in our current temporal pooling;
and (iii)~extend our proposed method to no-reference video quality assessment.
Finally, it is important to highlight that although our method achieves
impressive performance on the available 360 videos quality datasets, there is
still need in the immersive media community to produce more challenging
datasets, considering different distortions, projections, and HMD devices.
Overall, our proposed method can also be easily adaptable to new projection
types and other individual objective metrics that better maps the distortions
on specific 360 video processing contexts as well, which is also another
interesting future work direction.

\bibliographystyle{spbasic}
\bibliography{main}

\end{document}